\documentstyle[sprocl]{article}

\def\be{\begin{equation}}
\def\ee{\end{equation}}
\def\bea{\begin{eqnarray}}
\def\eea{\end{eqnarray}}

\begin{document}

\title{MASS HIERARCHIES, HIDDEN SYMMETRY AND MAXIMAL $CP$--VIOLATION\footnote[1]{Invited
Talk given at the ``6th International Symposium on Particles, Strings and
Cosmology (PASCOS--98), Northeastern University, Boston, MA, March 1998}}

\author{HARALD FRITZSCH}

\address{Universit\"at M\"unchen, Sektion Physik, Theresienstra\"se 37, \\
D--80333 M\"unchen\\E-mail: bm@hep.physik.uni--muenchen.de} 




\maketitle\abstracts{In view of the observed strong hierarchy of the quark and
lepton masses and
of the flavor mixing angles it is argued that the description of flavor
mixing must take this into account. One particular interesting way to describe
the flavor mixing, which, however, is not the one used today, emerges, which
is particularly suited for models of quark mass matrices based on flavor
symmetries. We conclude that the unitarity triangle important for $B$
physics should be close to or identical to a rectangular triangle. $CP$
violation is maximal in this sense.}

At the magnificient Boston Museum of Fine Arts one can see a big stone brought
in from Northern Africa, covered with strange hieroglyphes. More than 2000
years ago it located in the Great Temple of Amun at the old City of Jebel
Barkal in the kingdom of Nubia and is assumed to describe the rulership of
king Tanyidamani. The text is written in the Meroitic language, which is still
underdeciphered. Neither the grammar of that language nor the content of the
text on the Stone of Amun is known, only the letters.

In particle physics today one is facing a similar problem, as far as the
masses of the leptons and quarks are concerned. After the discovery of the
$t$--quark the spectrum of these masses (apart from the yet unknown neutrino
masses) is known. It is a rather wild spectrum, extending over 5 orders of
magnitude, from the tiny electron mass to the huge $t$--mass, but the actual
dynamics behind this spectrum remains mysterious. Nature speaks to us in some
kind of Meroitic language. The letters of this language, i. e. the masses and
flavor mixing parameters, are known, but the grammar and the content of the
text is unknown. Of course, in my talk I cannot offer a complete solution of
the mass problem, but I shall describe what I would like to define as the
grammar of patterns and rules, which are not only very simple, but seem to
come out very well, if confronted with the experimental results.

Let me remind you, just for illustration, of the observed eigenvalues of the
quark masses. Typical numbers are, at a renormalization point of
\[
\mu = m_t (\cong 175):
\]
\begin{equation}
\begin{array}{lclllclllclll}
m_u & : & 3.3 & MeV & \, m_c & : & 0.84 & GeV & \, m_t & : & 175 & GeV \\
m_d & : & 6.3 & MeV & \, m_s & : & 0.11 & GeV & \, m_b & : & 3.2 & GeV
\end{array}
\end{equation}
These masses are, of course, just eigenvalues of the quark mass
matrices, which in the Standard Model are introduced by the coupling of the
quark fields to the scalar field.

The phenomenon of flavor mixing arises due to the observed fact that the
$W$--boson, after interacting with a mass eigenstate, produces a state, which
is a mixture of all three quark mass eigenstates of the same electric charge.
Thus a $u$--quark, for example, is transformed primarily into a $d$--quark
(with a probability of about 95\%), sometimes into a $s$--quark (probability
about 5\%), and occasionally (probability about $10^{-5}$) into a $b$--quark,
provided that the energy transfer is large enough. This mismatch between the
$U$--sector and the $D$--sector \, of \, the quarks \, is usually
\, parametrized \, by the CKM mixing matrix \cite{{Cabibbo63},{KM73}}.



The phenomenon of flavor mixing, which is intrinsically linked to
$CP$--violation, is an important ingredient of the Standard Model of Basic
Interactions. Yet unlike other features of the Standard Model, e.\ g.\ the
mixing of the neutral electroweak gauge bosons, it is a phenomenon which can
merely be described. A deeper understanding is still lacking, but most
theoreticians would agree that it is directly linked to the mass spectrum of
the quarks -- the possible mixing of lepton flavors will not be discussed
here. Furthermore there is a general consensus that a deeper dynamical
understanding would require to go beyond the physics of the Standard Model.
In this talk I shall not go thus far. Instead I shall demonstrate that the
observed properties of the flavor mixing, combined with our knowledge about
the quark mass spectrum, suggest specific symmetry properties which allow
to fix the flavor mixing parameters with high precision, thus predicting the
outcome of the experiments which will soon be performed at the $B$--meson
factories.

In the standard electroweak theory the phenomenon of flavor
mixing of the quarks is described by the $3\times 3$ unitary CKM--matrix.
This
matrix can be expressed in terms of four parameters, which are usually 
taken as three rotation angles and one phase.

In the standard model the generation of quark masses 
is intimately related to the phenomenon of flavor mixing. In
particular, the flavor mixing parameters do depend on the
elements of quark mass matrices. A particular structure of the
underlying mass matrices calls for a particular choice of the
parametrization of the flavor mixing matrix. For example, in
ref. (3) it was noticed that a rather special form of
the flavor mixing matrix results, if one starts from Hermitian mass
matrices in which the (1,3) and (3,1) elements vanish. This has been
subsequently observed again in a number of papers
\cite{{Hall},{FX97}}. Recently we have studied the exact form of such a
description from a general point of view and pointed out some
advantages of this type of representation in the discussion of flavor
mixing and $CP$-violating phenomena \cite{FX97}, which will be discussed
later.

In the standard model the weak charged currents are given by 
\begin{equation}
\overline{(u, ~ c, ~ t)}^{~}_L \left ( \matrix{
V_{ud}  & V_{us}        & V_{ub} \cr
V_{cd}  & V_{cs}        & V_{cb} \cr 
V_{td}  & V_{ts}        & V_{tb} \cr} \right ) 
\left ( \matrix{
d \cr s \cr b \cr} \right  )_L \; ,
\end{equation}
where $u$, $c$, ..., $b$ are the quark mass eigenstates, $L$ denotes
the left-handed fields, and $V_{ij}$ are elements of the CKM matrix
$V$. In general $V_{ij}$ are complex numbers, but their absolute
values are measurable quantities. For example, $|V_{cb}|$ primarily
determines the lifetime of $B$ mesons. The phases of $V_{ij}$,
however, are not physical, like the phases of quark fields. A phase
transformation of the $u$ quark ($u \rightarrow u ~ e^{{\rm
i}\alpha}$), for example, leaves the quark mass term invariant but
changes the elements in the first row of $V$ (i.e., $V_{uj} \rightarrow 
V_{uj} ~ e^{-{\rm i}\alpha}$). Only a common phase transformation of all 
quark fields leaves all elements of $V$ invariant, thus there is a
five-fold freedom to adjust the phases of $V_{ij}$.

In general the unitary matrix $V$ depends on nine parameters.
Note that in the absence of complex phases $V$ would consist of only three 
independent parameters, corresponding to three (Euler) rotation
angles. Hence one can describe the complex matrix $V$ by three
angles and six phases. Due to the freedom in redefining the quark
field phases, five of the six phases in $V$ can be absorbed and we arrive
at the well-known result that the CKM matrix $V$ can be parametrized
in terms of three rotation angles and one $CP$-violating phase.

Recently it was shown that one way to describe the mixing of three families
is particularly useful. It is given as follows \cite{FX97}:
\begin{eqnarray}
V & = & \left( \matrix{
c_{\rm u}       & s_{\rm u}     & 0 \cr
-s_{\rm u}      & c_{\rm u}     & 0 \cr
0       & 0     & 1 \cr } \right )  \left ( \matrix{
e^{-i \varphi}       & 0     & 0 \cr
0       & c     & s \cr
0       & -s    & c \cr } \right )  \left ( \matrix{
c_{\rm d}       & -s_{\rm d}    & 0 \cr
s_{\rm d}       & c_{\rm d}     & 0 \cr
0       & 0     & 1 \cr } \right)  \nonumber \\ \nonumber \\ \nonumber
\end{eqnarray}
\vspace*{-1.5cm}
\begin{eqnarray}
= \left( \matrix{
s_u s_d c + c_u c_d e^{- i \varphi} & s_u c_d c - c_u s_d e^{-i \varphi}
& s_u s \cr
c_u s_d c - s_u c_d e^{-i \varphi} & c_u c_d c + s_u s_d e^{-i \varphi}
& c_u s \cr
- s_d s & - c_d s & c \cr } \right) .
\end{eqnarray}
The three angles $\theta_{\rm u}$, $\theta_{\rm d}$ and
$\theta$ in Eq. (12) can all be arranged to lie in the first quadrant
through a suitable redefinition of quark field phases. Consequently
all $s_{\rm u}$, $s_{\rm d}$, $s$ and $c_{\rm u}$, $c_{\rm d}$, $c$
are positive. The phase $\varphi$ can in general take values from 0
to $2\pi$; and $CP$ violation is present in weak interactions
if $\varphi \neq 0, \pi$ and $2\pi$.

In comparison with all other parametrizations discussed
previously \cite{KM73,Standard}, the one given here 
has a number of interesting features which in our view make it very
attractive and provide strong arguments for its use in future
discussions of flavor mixing phenomena, in particular, those in
$B$-meson physics. We shall discuss them below.

a) As shown in ref. (5), the flavor mixing matrix $V$ in Eq. (12) follows
directly from the chiral expansion of the mass
matrices. Thus it naturally takes into account the hierarchical structure of the 
quark mass spectrum.

b) The complex phase describing $CP$ violation ($\varphi$) appears only in the
(1,1), (1,2), (2,1) and (2,2) elements of $V$, i.e., 
in the elements involving only the quarks of the first and second
families. This is a natural description of $CP$ violation since in our 
hierarchical approach $CP$ violation is not directly linked to the third family, but
rather to the first and second ones, and in particular to the mass terms of the
$u$ and $d$ quarks. 

It is instructive to consider the special case $s_{\rm u} = s_{\rm d}
= s = 0$. Then the flavor mixing matrix $V$ takes the form
\begin{equation}
V \; = \; \left ( \matrix{
e^{-{\rm i}\varphi}     & 0     & 0 \cr
0       & 1     & 0 \cr
0       & 0     & 1 \cr} \right ) \; .
\end{equation}
This matrix describes a phase change in the weak transition between
$u$ and $d$, while no phase change is present in the
transitions between $c$ and $s$ as well as $t$ and $b$.
Of course, this effect can be absorbed in a phase change of the $u$-
and $d$-quark fields, and no $CP$ violation is present. Once the
angles $\theta_{\rm u}$, $\theta_{\rm d}$ and $\theta$ are introduced, 
however, $CP$ violation arises. It is due to a phase change in the weak
transition between $u^{\prime}$ and $d^{\prime}$, where $u^{\prime}$
and $d^{\prime}$ are the rotated quark fields, obtained by applying
the corresponding rotation matrices given in Eq. (12) to the 
quark mass eigenstates ($u^{\prime}$: mainly $u$, small admixture of
$c$; $d^{\prime}$: mainly $d$, small admixture of $s$).

c) The dynamics of flavor mixing can easily be interpreted by
considering certain limiting cases in Eq. (8). In the limit $\theta
\rightarrow 0$ (i.e., $s \rightarrow 0$ and $c\rightarrow 1$), the
flavor mixing is, of course, just a mixing between the first and
second families, described by only one mixing angle (the Cabibbo angle 
$\theta_{\rm C}$).  
It is a special and essential feature of the representation (8) that the Cabibbo
angle is {\it not} a basic angle, used in the parametrization. 
The matrix element $V_{us}$ (or $V_{cd}$) is
indeed a superposition of two terms including a phase. This feature
arises naturally in our hierarchical approach, but it is not new. In
many models of specific textures of mass matrices, it is indeed the
case that the Cabibbo-type transition $V_{us}$ (or $V_{cd}$) 
is a superposition of several
terms. At first, it was obtained by me 
in the discussion of the two-family mixing, and in various studies of quark
mass matrices \cite{Fritzsch79,Dimopoulos92}.

In the limit $\theta =0$ considered here, one has $|V_{us}| = |V_{cd}|
= \sin\theta_{\rm C} \equiv s^{~}_{\rm C}$ and
\begin{equation}
s^{~}_{\rm C} \; =\; \left | s_{\rm u} c_{\rm d} ~ - ~ c_{\rm u} s_{\rm d}
e^{-{\rm i}\varphi} \right | \; .
\end{equation}
This relation describes a triangle in the complex plane which we shall denote
as the ``LQ-- triangle'' (``light quark triangle''). This triangle is a
feature of the mixing of the first two families. Explicitly one has
(for $s=0$):
\begin{equation}
\tan\theta_{\rm C} \; =\; \sqrt{\frac{\tan^2\theta_{\rm u} +
\tan^2\theta_{\rm d} - 2 \tan\theta_{\rm u} \tan\theta_{\rm d}
\cos\varphi}
{1 + \tan^2\theta_{\rm u} \tan^2\theta_{\rm d} + 2 \tan\theta_{\rm u}
\tan\theta_{\rm d} \cos\varphi}} \; .
\end{equation}
Certainly the flavor mixing matrix $V$ cannot accommodate $CP$ violation in
this limit. However, the existence of $\varphi$ seems necessary in order
to make Eq. (6) compatible with current data, as one can see below.

d) The three mixing angles $\theta$, $\theta_{\rm u}$ and 
$\theta_{\rm d}$ have a precise physical meaning. The angle $\theta$
describes the mixing between the second and third families.
We shall refer to this mixing involving $t$ and $b$ as the ``heavy
quark mixing''.
The angle $\theta_{\rm u}$,
however, describes the $u$-$c$ mixing, and we shall denote this as the
``u-channel mixing''.
The angle $\theta_{\rm d}$ describes 
the $d$-$s$ mixing: it will be denoted as the ``d-channel mixing''. 
Thus there exists an asymmetry between the mixing of the first and
second families and that of the second and third families,
which in our view reflects interesting details of the underlying dynamics of
flavor mixing. 
The heavy quark mixing is a combined effect, involving both charge
$+2/3$ and charge $-1/3$ quarks, while the u- or d-channel mixing
(described by the angle $\theta_{\rm u}$ or $\theta_{\rm d}$) proceeds 
solely in the charge $+2/3$ or charge $-1/3$ sector. Therefore a precise
experimental determination of these two angles would allow to draw
interesting conclusions about the amount and perhaps the underlying
pattern of the u- or d-channel mixing.

e) The three angles $\theta$, $\theta_{\rm u}$ and $\theta_{\rm d}$
are related in a very simple way to observable quantities of $B$-meson 
physics. 
For example, $\theta$ is related to 
the rate of the semileptonic decay $B\rightarrow D^*l\nu^{~}_l$; 
$\theta_{\rm u}$ is associated with the ratio of the decay rate of
$B\rightarrow (\pi, \rho) l \nu^{~}_l$ to that of $B\rightarrow 
D^* l\nu^{~}_l$; and $\theta_{\rm d}$ can be determined from the ratio of
the mass difference between two $B_d$ mass eigenstates to that between
two $B_s$ mass eigenstates. We find the following exact
relations:
\begin{equation}
\sin \theta \; = \; |V_{cb}| \sqrt{ 1 + \left |\frac{V_{ub}}{V_{cb}}
\right |^2} \; ,
\end{equation}
and
\begin{eqnarray}
\tan\theta_{\rm u} & = & \left | \frac{V_{ub}}{V_{cb}} \right | \; ,
\nonumber \\
\tan\theta_{\rm d} & = & \left | \frac{V_{td}}{V_{ts}} \right | \; .
\end{eqnarray}
These simple results makes our parametrization (8) uniquely favorable 
for the study of $B$-meson physics.

By use of current data on $|V_{ub}|$ and $|V_{cb}|$, i.e., $|V_{cb}| = 
0.039 \pm 0.002$ \cite{Neubert96,Forty97} and $|V_{ub}/V_{cb}| =0.08 \pm 0.02$ 
\cite{PDG96}, we obtain $\theta_{\rm u} = 4.57^{\circ} \pm
1.14^{\circ}$ and $\theta = 2.25^{\circ} \pm 0.12^{\circ}$. Taking
$|V_{td}| = (8.6 \pm 2.1) \times 10^{-3}$,
which was obtained from the analysis of current data on
$B^0_d$-$\bar{B}^0_d$ mixing,
we get $|V_{td}/V_{ts}| = 0.22 \pm 0.07$, i.e., $\theta_{\rm d} = 12.7^{\circ} 
\pm 3.8^{\circ}$.
Both the heavy quark mixing angle $\theta$ and the u-channel mixing
angle $\theta_{\rm u}$ are relatively small. Recently a fit of these angles
was made \cite{PARO}, with rather small
uncertainties for the angles and the phase $\varphi $. One
finds:
\begin{eqnarray}
\Theta & = & (2.30 \pm 0.09)^0, \quad \Theta_u = (4.87 \pm 0.98)^0, \nonumber\\
\Theta_d & = & (11.71 \pm 1.09)^0, \quad \varphi = (91.1 \pm 11.8)^0
\end{eqnarray}
\\
These values are consistent with the ones given above, however the errors are
significantly smaller.

f) The phase $\varphi$ is
a phase difference between the contributions to $V_{us}$ (or $V_{cd}$) 
from the u-channel mixing and the d-channel mixing. The phase $\varphi$ is
not likely to be $0^{\circ}$ or $180^{\circ}$, according
to the experimental values given above, even though the measurement of 
$CP$ violation in $K^0$-$\bar{K}^0$ mixing is not taken
into account. For $\varphi =0^{\circ}$, one
finds $\tan\theta_{\rm C} = 0.14 \pm 0.08$; and for $\varphi =
180^{\circ}$, one gets $\tan\theta_{\rm C} = 0.30 \pm 0.08$. Both
cases are hardly consistent with the value of $\tan\theta_{\rm
C}$ obtained from experiments ($\tan\theta_{\rm C} \approx
|V_{us}/V_{ud}| \approx 0.226$). 

g) The $CP$-violating phase $\varphi$ in the flavor mixing matrix $V$ can be
determined from $|V_{us}|$ ($= 0.2205 \pm 0.0018$)
through the following formula, obtained easily from Eq. (12):
\begin{equation}
\varphi \; =\; \arccos \left ( \frac{s^2_{\rm u} c^2_{\rm d} c^2 +
c^2_{\rm u} s^2_{\rm d} - |V_{us}|^2}{2 s_{\rm u} c_{\rm u} s_{\rm d}
c_{\rm d} c} \right ) \; .
\end{equation}
\\
The two-fold ambiguity associated with the value of $\varphi$, coming
from $\cos\varphi = \cos (2\pi - \varphi)$, is removed if one
takes $\sin\varphi >0$ into account.
More precise measurements of the angles $\theta_{\rm u}$ and
$\theta_{\rm d}$ in the forthcoming experiments of $B$ physics will
remarkably reduce the uncertainty of $\varphi$ to be determined from Eq.
(10). This approach is of course complementary to the direct determination of
$\varphi$ from $CP$ asymmetries in some weak $B$-meson decays into hadronic
$CP$ eigenstates \cite{Sanda80}. 

Considering the presently known phenomenological constraints (see e.g.
ref. (7) that the value of $\varphi$ is most likely in the
range $40^{\circ}$ to $120^{\circ}$, the central value is
$\varphi \approx 81^{\circ}$. 
Note that $\varphi$ is essentially independent of the angle $\theta$,
due to the tiny observed value of the latter. 
Once $\tan\theta_{\rm d}$ is precisely measured, one shall be able to fix the
magnitude of $\varphi$ to a satisfactory degree of accuracy.\\
h) It is well--known that $CP$ violation in the flavor mixing matrix $V$ can
be described by the quantity ${\cal J}$ \cite{Jarlskog89}:
\begin{equation}
{\rm Im} \left( V_{il} V_{jm} V^*_{im} V^*_{jl} \right) = {\cal J}
\sum\limits^{3}_{k,n=1} \left( \epsilon_{ijk}\epsilon_{lmn} \right] \, .
\end{equation}
In our parametrisation ${\cal J}$ reads
\begin{equation}
{\cal J} = s_uc_us_dc_ds^2 \, \, c sin \varphi
\end{equation}
Obviously $\varphi = 90^{\circ }$ leads to the maximal value of ${\cal J}$.
Indeed $\varphi =90^{\circ}$, a particularly interesting case for $CP$ 
violation, is quite consistent with
current data. Since in our description of the flavor mixing the
complex phase $\varphi$ is related in a simple way to the phases of
the quark mass terms, the case $\varphi = 90^{\circ}$ is especially
interesting. It can hardly be an accident, and this case should be
studied further. The possibility that the phase $\varphi$ describing
$CP$ violation in the standard model is given by the algebraic number
$\pi/2$ should be taken seriously. It may provide a useful clue
towards a deeper understanding of the origin of $CP$ violation
and of the dynamical origin of the fermion masses, and might be a signed for
an interesting new symmetry (see also ref. (15)).

The case $\varphi =90^{\circ}$ has been
denoted as ``maximal'' $CP$ violation. It implies in our framework 
that in the complex
plane the u--channel and d--channel mixings are perpendicular to each
other. In this special case (as well as $\theta\rightarrow 0$), we have 
\begin{equation}
\tan^2\theta_{\rm C} \; =\; \frac{\tan^2\theta_{\rm u} ~ + ~
\tan^2\theta_{\rm d}}{1 ~ + ~ \tan^2\theta_{\rm u} \tan^2\theta_{\rm
d}} \; .
\end{equation}
To a good approximation (with the relative error $\sim 2\%$), 
one finds $s^2_{\rm C} \approx s^2_{\rm u} + s^2_{\rm d}$. 
h) At future $B$-meson factories, the study of $CP$ violation will
concentrate on measurements of the unitarity triangle 
\begin{equation}
S_u ~ + ~ S_c ~ + ~ S_t \; = \; 0 \; ,
\end{equation}
where $S_i \equiv V_{id} V^*_{ib}$ in the complex
plane. The inner angles of this triangle 
are as usual given by:
\begin{eqnarray}
\alpha & \equiv & \arg (- S_t S^*_u ) \; , \nonumber \\
\beta  & \equiv & \arg (- S_c S^*_t ) \; , \nonumber \\
\gamma & \equiv & \arg (- S_u S^*_c ) \; .
\end{eqnarray}
In terms of the parameters
$\theta$, $\theta_{\rm u}$, $\theta_{\rm d}$ and $\varphi$, we obtain
\begin{eqnarray}
\sin (2\alpha) & = & \frac{2 c_{\rm u} c_{\rm d} \sin\varphi \left
( s_{\rm u} s_{\rm d} c + c_{\rm u} c_{\rm d} \cos\varphi \right )}{s^2_{\rm
u} s^2_{\rm d} c^2 + c^2_{\rm u} c^2_{\rm d} + 2 s_{\rm u} c_{\rm u} s_{\rm d} c_{\rm d} c
\cos\varphi} \; , \nonumber \\ \nonumber \\
\sin (2\beta) & = & \frac{2 s_{\rm u} c_{\rm d} \sin\varphi \left
( c_{\rm u} s_{\rm d} c - s_{\rm u} c_{\rm d} \cos\varphi \right )}{c^2_{\rm
u} s^2_{\rm d} c^2 + s^2_{\rm u} c^2_{\rm d} - 2 s_{\rm u} c_{\rm u} s_{\rm d} c_{\rm d} c
\cos\varphi} \; .
\end{eqnarray}
To an excellent degree of accuracy, one finds $\alpha \approx
\varphi$. In order to illustrate how accurate this relation is, let us
use the central values of $\theta$, $\theta_{\rm u}$ and $\theta_{\rm 
d}$ (i.e., $\theta = 2.25^{\circ}$, $\theta_{\rm u} = 4.57^{\circ}$
and $\theta_{\rm d} = 12.7^{\circ}$). Then one arrives at
$\varphi - \alpha \approx 1^{\circ}$ as well as $\sin (2\alpha)
\approx 0.34$ and $\sin (2\beta) \approx 0.65$. 
It is expected that $\sin (2\alpha)$ and $\sin (2\beta)$
will be directly measured from the $CP$ asymmetries in 
$B_d \rightarrow \pi^+\pi^-$ and $B_d \rightarrow J /\psi K_S$ modes
at a $B$-meson factory.

Note that the three sides of the unitarity triangle 
can be rescaled by $|V_{cb}|$. In a very good approximation
(with the relative error $\sim 2\%$), one arrives at
\begin{equation}
|S_u| ~ : ~ |S_c| ~ : ~ |S_t| \; \approx \; s_{\rm u} c_{\rm d} ~ : ~ 
s^{~}_{\rm C} ~ : ~ s_{\rm d} \; .
\end{equation}
Equivalently, one can obtain
\begin{equation}
s_{\alpha} ~ : ~ s^{~}_{\beta} ~ : ~ s_{\gamma} \; \approx \; s^{~}_{\rm C} 
~ : ~ s_{\rm u} c_{\rm d} ~ : ~ s_{\rm d} \; ,
\end{equation}
where $s_{\alpha} \equiv \sin\alpha$, etc.
Comparing this triangle with the LQ--triangle we find that they are 
indeed congruent with each other to a high degree of accuracy.
The congruent relation between these two triangles is particularly
interesting, since the LQ--triangle is essentially a feature of the physics
of the first two quark families, while the unitarity triangle by
defination is
linked to all three families. In this connection it is of special
interest to note that in models which specify the textures of the mass 
matrices the Cabibbo triangle and hence the three angles of the unitarity
triangle can be fixed by the spectrum of the light quark masses and
the $CP$-violating phase $\varphi$.\\
j) Compared with the standard parametrization of the flavor mixing
matrix $V$ the parametrization discussed here has an additional
advantage: the renormalization-group evolution of $V$, from the weak
scale to an arbitrary high energy scale, is 
to a very good approximation associated only with the angle $\theta$. This can
easily be seen if one keeps the $t$ and $b$ Yukawa couplings only  
and neglects possible threshold effect in the one-loop
renormalization-group equations of the Yukawa matrices \cite{RGE}.
Thus the parameters $\theta_{\rm u}$, $\theta_{\rm d}$ and $\varphi$
are essentially independent of the energy scale, while $\theta$ does
depend on it and will change if the underlying scale is shifted, say
from the weak scale ($\sim 10^2$ GeV) to the grand unified theory
scale (of order $ 10^{16}$ GeV). In short, the heavy quark mixing is
subject to renormalization-group effects; but the u- and d-channel
mixings are not, likewise the phase $\varphi$ describing $CP$
violation and the LQ--triangle as a whole. It follows that only the angle
$\theta $, but not $\Theta _u, \Theta _d$ or $\varphi $, depends in its
behaviour on the reference energy scale and is increased on the underlying
model, e.\ g.\ on whether there is a supersymmetric extension of the Standard
Model or not.

We have presented a new description of the flavor mixing 
phenomenon, which is based on the phenomenological fact that the quark 
mass spectrum exhibits a clear hierarchy pattern. This leads uniquely
to the interpretation of the flavor mixing in terms of a heavy quark
mixing, followed by the u-channel and d-channel mixings. The complex
phase $\varphi$, describing the relative orientation of the u-channel
mixing and the d-channel mixing in the complex plane, signifies
$CP$ violation, which is a phenomenon primarily linked to the physics
of the first two families. The Cabibbo angle is not a basic mixing
parameter, but given by a superposition of two terms involving the
complex phase $\varphi$. The experimental data suggest that the phase
$\varphi$, which is directly linked to the phases of the quark mass
terms, is close to $90^{\circ}$. This opens the possibility to
interpret $CP$ violation as a maximal effect, in a similar way as
parity violation. 

Our description of flavor mixing has many clear advantages compared
with other descriptions. We propose that it should be used in the
future description of flavor mixing and $CP$ violation, in particular, 
for the studies of quark mass matrices and $B$-meson physics.

The description of the flavor mixing phenomenon given above is of special
interest if for the $U$ and $D$ channel mixing the quark mass textures
discussed first in \cite{Fritzsch77} are applied (see also \cite{FX98}).
In that case one finds \cite{RGE} (apart from small corrections)
\begin{equation}
{\rm tan} \Theta _d = \sqrt{\frac{m_d}{m_s}}
\end{equation}
\[
{\rm tan} \Theta _u = \sqrt{\frac{m_u}{m_c}} \, .
\]
The experimental value for ${\rm tan} \, \Theta _u$ given by the ratio
$V_{ub} / V_{cb} $ is in agreement with the observed value for
$\left( m_u / m_c \right) ^{1/2} \approx 0.07$, but the errors for both
$\left( m_u / m_c \right) ^{1/2}$ and $V_{ub} / V_{cb}$ are comparable
(about 25\%).

The angle $\Theta_d $ is expected to be about 12.6$^{\circ}$, if we use a
mass ratio $m_s / m_d \approx 20$, as obtained in chiral perturbation theory.
This agrees well with the experimental values discussed above.

As emphasized in ref.\ (17), the phase angle $\varphi $ is very close
to 90$^{\circ }$, implying that the LQ--triangle and the
unitarity triangle are essentially rectangular triangles. In particular the
angle $\beta $ which is likely to be measured soon in the study of the
reaction $B^{\circ } \rightarrow J / \psi K^{\circ }_s$ is expected to be close
to $20 ^{\circ }$.

It will be very interesting to see whether the angles $\Theta _d$ and
$\Theta _u$ are indeed given by the square roots of the light quark mass
ration $m_d / m_s$ and $m_u /m_c$, which imply that the phase $\varphi$
is close to or exactly $90 ^{\circ }$. This would mean that the light quarks
play the most important r$\hat{\rm o}$le in the dynamics of flavor mixing and
$CP$ violation.\\
\\



%
%
%


\end{document}